\documentclass[pre,twocolumn,showpacs]{revtex4}
\usepackage{bm}
\usepackage{amssymb}
\begin{document}
\title{Self interaction near dielectrics}
\author{Lior M.~Burko}
\affiliation{Department of Physics, University of Utah, Salt Lake
City, Utah 84112}
\date{\today}
\begin{abstract} 
We compute the force acting on a free, static electric charge outside a
uniform dielectric sphere. We view this force as a self-interaction force,
and compute it by applying the Lorentz force directly to the charge's
electric field at its location. We regularize the divergent bare
force using two different methods: A direct application of the Quinn-Wald
Comparison Axiom, and Mode-Sum Regularization.
\end{abstract}
\maketitle

\section{Introduction}

Electric charges which are placed in an inhomogeneous ponderable medium 
undergo self interaction. The simplest case is that of a static electric
charge  in an inhomogeneous dielectric. The self interaction of the
charge results in a self force (in other contexts this self force is also
known as a radiation reaction force), which acts to accelerate the
charge. In the static problem, then, one can ask the following
question: What is the external, possibly non-electric, force which needs to
be exerted on the charge to keep it static when a non-uniform dielectric is
present? 

In this paper we study this question for a very simple case. Specifically,
we find the self force on a pointlike electric charge $e$ outside a uniform
dielectric sphere. (By an inhomogeneous dielectric we here mean the
discontinuity of the dielectric constant at the surface of the sphere.) The
origin of the force on the charge in this case is simple: The charge
polarizes the dielectric at order $e$. The induced electric field then
back-reacts on the original charge, and this interaction then is at order
$e^2$. In fact, one can compute the force on the charge following this
simple physical picture. However, one can employ a different picture, in
which the force on the charge is construed as a self force. The charge
interacts with its own field, and the latter is distorted by the presence
of the dielectric sphere. In this picture the force is computed {\it locally}, 
using only the fields at the location of the charge. The local
approach has many merits. Specifically, the computation of the near field
is much simpler, and one does not have to compute additional quantities
such as the far field or the sphere's polarization. 
(The fields at great distances of course contribute to the force on the
charge, but for
the local approach only through boundary conditions.) A similar approach
was used in Ref.\ \cite{ajp} to compute the radiated power of synchrotron
radiation using only the near field. The difficulty in
the local approach arises from the well known fact that the field of a
point charge diverges when the evaluation point for the field coincides
with the field's source. In fact, this happens already for a static charge
in empty (and flat) spacetime. Let the charge $e$ be located on the 
${\bf {\hat z}}$-axis in spherical coordinates. Decomposing the scalar
potential
$\Phi$ into Legendre polynomials, one finds that 
$$\Phi(r,\vartheta)=e\sum_{\ell=0}^{\infty}\frac{r_{<}^{\ell}}{r_{>}^{\ell+1}}
P_{\ell}(\cos\vartheta),$$
where $r_<\, (r_>)$ is the smaller (greater) of the $r$ values of the 
source's location $r_0$ and the evaluation point. The (self) force acting
on the
charge is then given by the (average of the two one-sided) gradients of
the potential. Specifically,
$${\bf f}=-\frac{1}{2}e\left.\left(
{\bf \nabla}\Phi^++{\bf \nabla}\Phi^-\right)\right|_{{\bf r}=r_{0}
{\bf {\hat z}}},$$
or
$$f_r=\sum_{\ell=0}^{\infty}\frac{e^2}{2r_0^2}$$
which clearly diverges. (The derivation of the last equation is given 
below.) 
In this illustration of the problem, of course, it
is clear that the regularized, physical self force vanishes: The force
obviously cannot depend on where we choose to put the origin of our
coordinate system. (Also, we have ample observational evidence that
static isolated charges remain static.) 

There is a long history of works
on the self force. (For reviews see, e.g., \cite{reviews}.) Recently, the
analogous problem of calculation of self forces in curved spacetimes (also
for the gravitational case where the self interaction pushes a body with
finite mass off a geodesic) has gained much interest \cite{sf,msr}. In
this paper
we shall make use of some of the techniques, which have been developed for
self interaction in curved spacetime, for the problem of
interest. (Interestingly, there is a close link between electromagnetism
in static gravitational fields and electromagnetism in matter. As is well
known \cite{Moller}, Maxwell's equations in vacuum in static curved
spacetime
can be written as Maxwell's equations in flat spacetime with an effective
non-uniform dielectric.) Specifically, we shall make use of the Quinn-Wald
Comparison Axiom \cite{quinn-wald} and Mode-Sum Regularization \cite{msr} 
in order to extract the physical, finite piece of the self force.

The organization of this paper is as follows. In Section \ref{derivation}
we solve for the scalar potential, and obtain the modes of the bare 
force. This is, in fact, a standard exercise
in electromagnetism \cite{panofsky-phillips}. Then, in Section
\ref{regularize} we regularize the
self force using two different approaches, and in Section \ref{prop} we
discuss the properties of our result.

\section{Derivation of the bare force}\label{derivation}

\begin{figure}
\input epsf
\centerline{ \epsfxsize 7.0cm
\epsfbox{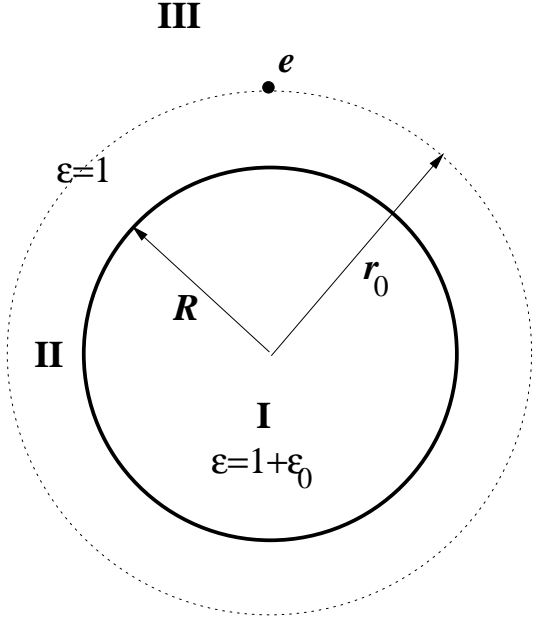}}
\caption{An electric charge $e$ at a distance $r_0$ from the center of a
sphere of radius $R$. The sphere has a dielectric constant
$\epsilon=1+\epsilon_0$, and outside the sphere the dielectric constant is
unity. Region I is for $r<R$, region II for $R<r<r_0$, and region III for
$r>r_0$.
}
\label{fig1}
\end{figure}

Consider a static electric charge $e$ in vacuum at radius $r_0$, outside
an insulated sphere of radius $R$ of uniform dielectric constant
$\epsilon=1+\epsilon_0$, where $\epsilon_0>0$. Notice, that $\epsilon_0$ 
is not the permittivity of free space, but rather 
$\epsilon_0=4\pi\chi_e$, where $\chi_e$ is the electric susceptibility. 
We place the charge $e$ on the ${\bf {\hat z}}$-axis without loss of
generality.  This configuration is plotted in Figure \ref{fig1}.

Maxwell's equation in
matter is 
\begin{equation}
{\bf \nabla \cdot D}=4\pi\rho
\label{max}
\end{equation}
where ${\bf D}=\epsilon {\bf E}$ is the displacement field, ${\bf E}$ is
the electric field, and $\rho$ is the density of free charges.

We assume that the distribution of the dielectric is spherically
symmetric (although non-uniform). Specifically, we take $\epsilon=\epsilon
(r)$.  (Despite the uniformity of the sphere, the dielectric constant
throughout space depends on $r$: it suffers a step-function discontinuity
at the surface of the sphere.) In the usual spherical coordinates Eq.\
(\ref{max}) becomes \begin{eqnarray}
\partial_{r}^2\Phi &+& 
\left(\frac{2}{r}+\frac{\partial_r\epsilon}{\epsilon}\right)
\partial_r\Phi+\frac{1}{r^2\sin\vartheta}\partial_{\vartheta}
\left(\sin\vartheta\, 
\partial_{\vartheta}\Phi\right)\nonumber \\
&=& -\frac{4\pi }{\epsilon (r)}\rho
\label{eq1}
\end{eqnarray}
where $\Phi$ is the scalar potential. 
We next decompose Eq.\ (\ref{eq1}) into Legendre polynomials. That is, 
$\Phi(r,\vartheta)=\sum_{\ell}\phi^{\ell}(r)P_{\ell}(\cos\vartheta)$ and 
$\rho(r,\vartheta)=\frac{e}{4\pi}\frac{\delta
(r-r_0)}{r_0^2}\sum_{\ell}
(2\ell +1)P_{\ell}(\cos\vartheta)$. 
The radial equation then becomes
\begin{eqnarray}
\partial_{r}^2\phi^{\ell} &+&
\left(\frac{2}{r}+\frac{\partial_r\epsilon}{\epsilon}\right)
\partial_r\phi^{\ell}-\frac{\ell (\ell+1)}{r^2}\phi^{\ell}
\nonumber \\
&=&-(2\ell +1)e\frac{\delta (r-r_0)}{\epsilon (r_0)r_0^2}\, .
\label{rad-eq}
\end{eqnarray}
The boundary conditions for this equation are that $\phi^{\ell}$ is
continuous everywhere (which includes regularity at the origin and at
infinity), but $\partial_r\phi^{\ell}$ is discontinuous at $r=R$ and at
$r=r_0$. Specifically, these latter two conditions are that 
$$\lim_{\sigma\to 0^+}\epsilon (R-\sigma)\partial_r\phi^{\ell}(R-\sigma)=
\lim_{\sigma\to 0^+}\epsilon (R+\sigma)\partial_r\phi^{\ell}(R+\sigma)$$
(which comes from the continuity of the normal component of the
displacement field at the surface of discontinuity), and 
$$\lim_{\sigma\to 0^+}\left[\partial_r\phi^{\ell}(r_0+\sigma)
-\partial_r\phi^{\ell}(r_0-\sigma)\right]=-(2\ell +1)\frac{e}{r_0^2},$$
[which comes from integration of Eq.\ (\ref{rad-eq}) across $r=r_0$, and
using the continuity of $\phi^{\ell}$ and $\epsilon$ there (the only
discontinuity of $\epsilon$ is at $r=R$)]. 

The radial functions $\phi^{\ell}$ then satisfy
\begin{eqnarray}
\phi^{\ell}(r)=\left \{ \begin{array}{lll}
A_{\ell}r^{\ell} & r<R & ({\rm region \;\; I}) \\
B_{\ell}r^{\ell}+C_{\ell}r^{-\ell -1} & R<r<r_0 & ({\rm region
\;\; II}) \\
D_{\ell}r^{-\ell -1} & r>r_0 & ({\rm region \;\; II})
\end{array} \right. \, ,
\end{eqnarray}
where the coefficients $A_{\ell},B_{\ell},C_{\ell}$ and $D_{\ell}$ are
found from the boundary conditions. We find that 
\begin{eqnarray}
A_{\ell}&=& \frac{2\ell +1}{2\ell +1+\ell\epsilon_0}\frac{e}{r_0^{\ell+1}}\\
B_{\ell}&=& \frac{e}{r_0^{\ell+1}}\\
C_{\ell}&=& -\frac{\ell}{2\ell+1+\ell\epsilon_0}
\frac{R^{2\ell+1}}{r_0^{\ell+1}}e\epsilon_0\\
D_{\ell}&=& e r_0^{\ell}\left[1-\frac{\ell}{2\ell +1+\ell\epsilon_0}
\left(\frac{R}{r_0}\right)^{2\ell+1}\epsilon_0\right]\, ,
\end{eqnarray}
such that the scalar potential $\Phi$ is given by
\begin{eqnarray}
\Phi=
\left \{ \begin{array}{ll}
e\sum\limits_{\ell=0}^{\infty}\frac{2\ell+1}{2\ell+1+\ell\epsilon_0}
\frac{r^{\ell}}{r_0^{\ell+1}}P_{\ell}(\cos\vartheta) & r<R \\
\Phi_{\rm vac}-\sum\limits_{\ell=0}^{\infty}
\frac{\ell}{2\ell+1+\ell\epsilon_0}\frac{R^{2\ell+1}}{r_0^{\ell+1}}
\frac{e\epsilon_0}{r^{\ell+1}}P_{\ell}(\cos\vartheta) & r>R
\end{array} \right. \, .
\label{potential}
\end{eqnarray}
Here, 
\begin{eqnarray}
\Phi_{\rm vac}&\equiv& \frac{e}{ \left| {\bf r}-r_0{\bf {\hat z}}
\right|}=e
\sum\limits_{\ell}\frac{r_<^{\ell}}{r_>^{\ell+1}}P_{\ell}(\cos\vartheta)
\label{potential1}
\end{eqnarray} 
is the potential in the absence of a dielectric sphere. 

The bare force ${\bf f}^{\rm bare}$ is found by 
${\bf f}^{\rm bare}=-e{\bf \nabla}\Phi$, evaluated at the location 
of the charge at $r=r_0$ and $\vartheta=0$. From symmetry, it is clear
that any force is radial. We compute, then, the radial component of the
force only. Differentiating Eq.\
(\ref{potential}) and using Eq.\ (\ref{potential1}), we find that
\begin{eqnarray}
f_r^{\rm bare}&=&\sum_{\ell=0}^{\infty}f_r^{\ell}\nonumber \\
&=& 
\sum_{\ell=0}^{\infty}
\left[\frac{e^2}{2r_0^2}-\frac{\ell(\ell+1)}{2\ell+1+\ell\epsilon_0}
\left(\frac{R}{r_0}\right)^{2\ell+1}\frac{\epsilon_0 e^2}{r_0^2}\right] \,
, \label{bare}
\end{eqnarray}
where $f^{\ell}_r=-\frac{e}{2}\lim_{\sigma\to 0^+}
[\partial_r\phi^{\ell}(r_0+\sigma)+\partial_r\phi^{\ell}(r_0-\sigma)]$.
Clearly, Eq.\ (\ref{bare}) diverges. This comes as no surprise, as we have
already mentioned that this divergence occures already for a charge in
empty space. In the next section we shall extract the physical, finite
part of this infinite bare force.

\section{Regularization of the bare force}\label{regularize}

In order to regularize the bare force (\ref{bare}), we make direct use of
the Quinn-Wald comparison axiom, for which plausible arguments were
given. The 
Comparison Axiom states the following (see \cite{quinn-wald} for more
details):
{\it Consider two points, $P$ and
$\tilde{P}$, each lying on time-like world lines in possibly different
spacetimes which
contain Maxwell fields $F_{\mu\nu}$ and ${\tilde F}_{\mu\nu}$ sourced by
particles of charge $e$ on
the world lines. If the four-accelerations of the world lines at $P$ and
$\tilde{P}$ have the
same magnitude, and if we identify the neighborhoods of $P$ and ${\tilde
P}$ via the
exponential map such that the four-velocities and four-accelerations are
identified
via Riemann normal coordinates, then the difference between the
electromagnetic forces
$f_{\mu}$ and ${\tilde f}_{\mu}$ is given by the
limit $x\to 0$ of the Lorentz force associated with the difference of the
two fields averaged over a sphere at geodesic distance $x$ from the world
line at $P$, i.e.,}
\begin{equation}
f_{\mu}-{\tilde f}_{\mu}=\lim_{x\to 0} 
e\left<F_{\mu\nu}-{\tilde F}_{\mu\nu}\right>_x u^{\nu}\ .
\label{sc:ax}
\end{equation}

Here, we identify the ``tilde'' spacetime as that of a globally empty
spacetime. Obviously, ${\tilde f}_{\mu}=0$.  
We emphasize that this axiom assumes a nearly trivial form for the case of
interest: The local neighborhood of the particle in question and of a
similar particle in a (globally-)empty spacetime are identical. (It is
only the far-away properties of spacetime -- as represented by different
dielectric constants -- which are different for the two spacetimes.) 
Another remark is that we do not need to average here over directions, as
the forces in our case are direction independent. Consider now Eq.\
(\ref{potential}) for the potential. Outside the dielectric sphere the
potential $\Phi$ contains the vacuum potential $\Phi_{\rm vac}$ and a
correction $\Delta\Phi$. We next use $\Phi_{\rm vac}$ to construct the
fields ${\tilde F}_{\mu\nu}$. Applying the Comparison Axiom, we find that
the self force is given by 
\begin{eqnarray}
f_r&=& -\sum_{\ell=0}^{\infty}
\frac{\ell(\ell+1)}{2\ell+1+\ell\epsilon_0}   
\left(\frac{R}{r_0}\right)^{2\ell+1}\frac{\epsilon_0 e^2}{r_0^2} 
\label{reg}
\\
&=& -\frac{2}{3+\epsilon_0}\left(\frac{R}{r_0}\right)^3\nonumber 
\\
&\times &
{{_2}F_{1}}\left[3,\frac{3+\epsilon_0}{2+\epsilon_0};\frac{5+2\epsilon_0}
{2+\epsilon_0};\left(\frac{R}{r_0}\right)^2\right]\frac{\epsilon_0 e^2}
{r_0^2}
\label{result}
\, , \end{eqnarray}
${{_2}F_{1}}$ being the hypergeometric function. We were unable to find
this result in the literature. (In view of the vastness of the literature
on classical electromagnetism, our search in the literature is naturally
incomplete.) 

Before we analyze the properties of this result, let us derive it using a
second method. Specifically, we use Mode-Sum
Regularization. (Note, that Mode-Sum Regularization is based on 
the Quinn-Wald result for the self force in curved spacetime, the latter
being a consequence of the Comparison Axiom. In that sense, these two
methods are not entirely independent. Here, however, we make direct use of
the Comparison Axiom, which is necessary but not sufficient in order to
derive the Quinn-Wald result.) 
Mode-Sum Regularization is
described in Refs.\ \cite{msr}. In Mode-Sum regularization one finds two
regularization functions, $h^{\ell}_{\mu}$ and $d_{\mu}$. 
The regularized self force is given by 
\begin{equation}
f_{\mu}=\sum_{\ell=0}^{\infty}\left(f_{\mu}^{\ell\; {\rm bare}}-
h^{\ell}_{\mu}\right)-d_{\mu}
\end{equation}
where $d_{\mu}$ is a finite valued function and $h^{\ell}_{\mu}$ has the
general form $h^{\ell}_r=a_r(\ell+\frac{1}{2})+b_r+c_r
(\ell+\frac{1}{2})^{-1}$. One only
needs the {\it local} properties of spacetime in order to determine these
functions. As locally the charge is in empty space (it is removed from the 
dielectric sphere), it is clear that the regularization functions 
$h^{\ell}_{\mu}$ and $d_{\mu}$ would be the same as in a
globally-empty spacetime. Indeed, it is easy to find the limit
as $\ell\to\infty$ of the modes of the bare force. The 
modes of the  radial component of the bare force (\ref{bare}) approach
$e^2/(2r_0^2)$ as $\ell\to\infty$. As $h^{\ell}_{\mu}$ must have the same
asymptotic structure (as $\ell\to\infty$) as $f^{\ell}_{\mu}$, this
implies that $h^{\ell}_{r}=e^2/(2r_0^2)$, identically the same as in 
(globally-)empty spacetime, in agreement with the previous reasoning. 
We similarly expect the function $d_r$ to vanish, as it does in a
globally-empty flat spacetime. We justify this expectation {\it a
posteriori} by demonstrating that this leads to the same expression as we
received by using the Comparison Axiom. It then follows that the
regularized self force is given by 
$f_r=\sum\limits_{\ell}[f_r^{\ell\, {\rm bare}}-e^2/(2r_0^2)]$, which
agrees with 
Eq.\ (\ref{reg}). 

\section{Properties of the result}\label{prop}

We found that the self force on the charge $e$ is given by Eq.\
(\ref{result}). This is an attractive force, as indeed is expected. (The
charge $e$ polarizes the sphere such that there is an excess of 
oppositely-charged induced charge on the sphere closer to the free
charge. Hence the polarization charge acts to attract the free charge.) We
can check our result in the limiting case of infinite dielectric,
$\epsilon_0\to\infty$,
which corresponds to the case of an uncharged, insulated, conducting
sphere. In that limit our result becomes
\begin{eqnarray}
f_r\rightarrow
&-&2\left(\frac{R}{r_0}\right)^3{_{2}F_{1}}\left[3,1;2;\left(\frac{R}{r_0}
\right)^2\right]\frac{e^2}{r_0^2}\nonumber \\
&=&-\frac{2r_0^2-R^2}{(r_0^2-R^2)^2}\left(\frac{R}{r_0}\right)^3e^2
\label{cond}
\end{eqnarray}
which is indeed the known result for an uncharged, insulated, conducting
sphere \cite{jackson}. The opposite extreme case is the limit as
$\epsilon_0\to 0$. Linearizing our result in $\epsilon_0$, we find that 
\begin{eqnarray}
f_r=&-&\frac{\sqrt{\pi}}{2}\left(\frac{R}{r_0}\right)^{3/2}
\left[1-\left(\frac{R}{r_0}\right)^2\right]^{-3/2}
\nonumber \\
&\times&
P_{1/2}^{-3/2}
\left(\frac{r_0^2+R^2}{r_0^2-R^2}\right)\frac{\epsilon_0 e^2}{r_0^2}
+O\left(\epsilon_0^2\right)
\, ,
\label{lin}
\end{eqnarray}
which vanishes linearly with $\epsilon_0$ as $\epsilon_0\to 0$. 

\begin{figure}
\input epsf
\centerline{ \epsfxsize 8.5cm
\epsfbox{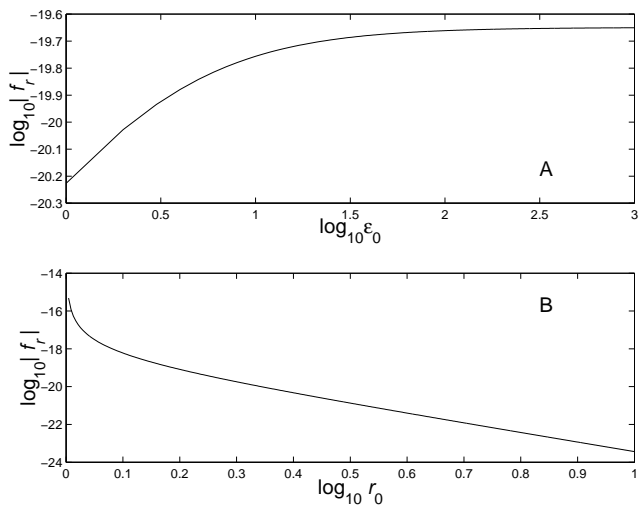}}
\caption{The self force on a free charge outside a dielectric sphere.
The charge $e$ is taken to be that of an electron, and the radius of the
sphere is $R=1{\rm cm}$.
Upper panel (A): The self force as a function of $\epsilon_0$, for
$r_0=2{\rm cm}$.
Lower panel (B): The self force as a function of $r_0$ (in cm), for
$\epsilon_0=10.7$.
}
\label{fig2}
\end{figure}

For any finite value of $\epsilon_0$ the force is smaller in
magnitude than in the case of a conducting sphere (\ref{cond}). This
behavior is shown in Fig.\ \ref{fig2}A, which plots 
the self force as a function of $\epsilon_0$ for fixed $r_0$. It can be
seen that as $\epsilon_0\to\infty$, the full expression approaches the
saturation value of the conducting sphere. 

At very large distances ($r_0 \gg R$), the self force becomes 
\begin{equation}
f_r=-\frac{2}{3+\epsilon_0}\left(\frac{R}{r_0}\right)^3
\frac{\epsilon_0 e^2}{r_0^2}+O(r_0^{-7})\, ,
\label{large_r}
\end{equation}
which drops off like $r_0^{-5}$. This behavior 
can be seen from Fig.\ \ref{fig2}B, which displays the self force 
as a function of $r_0$ for fixed $\epsilon_0$. We can check the validity of
this limit by deriving Eq.~(\ref{large_r}) using the alternative picture
\cite{ref}. When $r_0\gg R$, we can treat the field of the
charge $e$ to the leading order in $1/r_0$ as a constant over the
sphere. Let us take for simplicity the charge $e={\rm sgn} (e)|e|$ to be on
the positive ${\bf {\hat z}}$-axis. The electric field due to the free
charge is ${\bf E}_0=- {\rm sgn} (e) |e|r_0^{-2} {\bf {\hat z}}$, and the
polarization of the sphere then is just a constant inside the sphere,
and is given by 
${\bf P}=[3/(4\pi)][\epsilon_0/(3+\epsilon_0)]{\bf E}_0$. The 
dipole moment ${\bf p}$ can be obtained by a volume integral over ${\bf
P}$. One finds then that ${\bf p}=[\epsilon_0/(3+\epsilon_0)]R^3{\bf
E}_0$.  The electric field ${\bf E}$ at $r_0{\bf {\hat z}}$ is
found by ${\bf E}=[3{\bf {\hat z}}({\bf p}\cdot{\bf {\hat z}})-{\bf
p}]/r_0^3$, or 
${\bf E}=-2pr_0^{-3}{\rm sgn} (e){\bf {\hat z}}$, where 
$p\equiv|{\bf p}|=\epsilon_0R^3r_0^{-2}|e|/(3+\epsilon_0)$. The
force on the charge $e$ is simply ${\bf f}=e{\bf E}=-2pr_0^{-3}|e|{\bf
{\hat z}}$, which is equal to the
leading order term of Eq.~(\ref{large_r}). 

When $r_0$ approaches $R$ the self force grows rapidly, and in the limit
diverges. This is indeed expected: in this limit one has a point charge
near a semi-infinite dielectric. The solution for the force is a
classic image problem \cite{jackson}, which obviously diverges in the
coincidence limit of the charge and its image. This divergence happens
already in the case of the conducting sphere, as is evident from Eq.\
(\ref{cond}).  In fact, we find that the self force diverges whenever the
free charge is locally at a region with non-zero gradient of the
dielectric constant. 

We note that the magnitude of this self force is not extremely small for
realistic parameters. Take the charge $e$ to be that of an electron of
mass $m_e$, and the dielectric sphere to be made of Silicon, for which
$\epsilon_0=10.7$ at room temperature and pressure, and take the sphere to
be of radius $1{\rm cm}$. In the  gravitational field of the Earth, with
gravitational acceleration of $980{\rm cm}/{\rm sec}^2$, the self force
equals the weight of the electron when $r_0=13.2{\rm cm}$

\section*{Acknowledgments}
I thank Richard Price for discussions. This research was supported by the  
National Science Foundation through grant No.\ PHY-9734871.


\begin{thebibliography}{}
\bibitem{ref} We are thankful to an anonymous referee for suggesting this
check to us.
\end{thebibliography}
\end{document}